\documentclass[aps,pre,twocolumn,superscriptaddress,nofootinbib]{revtex4}

\usepackage{amsmath,amssymb}
\usepackage{dsfont}
\usepackage{bm}
\usepackage[dvipdfmx]{graphicx}
\usepackage[caption=false]{subfig}

\usepackage{color}

\begin{document}

\title{A density functional approach to ferrogels}

\author{P.\ Cremer}
\email{pcremer@thphy.uni-duesseldorf.de}
\affiliation{Institut f\"ur Theoretische Physik II: Weiche Materie, Heinrich-Heine-Universit\"at D\"usseldorf, D-40225 D\"usseldorf, Germany}
\author{M.\ Heinen}
\affiliation{Departamento de Ingenier\'{i}a F\'{i}sica, Divisi\'{o}n de Ciencias e Ingenier\'{i}as, University of Guanajuato, Loma del Bosque 103, 37150 Le\'{o}n, M\'{e}xico}
\author{A. M.\ Menzel}
\affiliation{Institut f\"ur Theoretische Physik II: Weiche Materie, Heinrich-Heine-Universit\"at D\"usseldorf, D-40225 D\"usseldorf, Germany}
\author{H.\ L\"owen}
\affiliation{Institut f\"ur Theoretische Physik II: Weiche Materie, Heinrich-Heine-Universit\"at D\"usseldorf, D-40225 D\"usseldorf, Germany}
\date{\today}
%

\begin{abstract}
Ferrogels consist of magnetic colloidal particles embedded in an elastic polymer matrix. 
As a consequence, their structural and rheological properties are governed by a competition between magnetic particle--particle interactions and mechanical matrix elasticity. 
Typically, the particles are permanently fixed within the matrix, which makes them distinguishable by their positions. 
Over time, particle neighbors do not change due to the fixation by the matrix. 
Here we present a classical density functional approach for such ferrogels. 
We map the elastic matrix-induced interactions between neighboring colloidal particles distinguishable by their positions onto effective pairwise interactions between indistinguishable particles similar to  a ``pairwise pseudopotential''.
Using Monte-Carlo computer simulations, we demonstrate for one-dimensional dipole-spring models of ferrogels that this mapping is justified.
We then use the pseudopotential as an input into classical density functional theory of inhomogeneous fluids and predict the bulk elastic modulus of the ferrogel under various conditions. 
In addition, we propose the use of an ``external pseudopotential'' when one switches from the viewpoint of a one-dimensional dipole-spring object to a one-dimensional chain embedded in an infinitely extended bulk matrix. 
Our mapping approach paves the way to describe various inhomogeneous situations of ferrogels using classical density functional concepts of inhomogeneous fluids. 
\end{abstract}

\pacs{}















\maketitle


\section{Introduction}
\label{Sec.Introduction}
Classical density functional theory for inhomogeneous fluids is nowadays used for many-body systems governed by a pair potential  (such as hard or soft spheres) and has found widespread applications for phase separation, freezing and interfacial phenomena, for reviews see Refs.~\onlinecite{Evans2016_JPhysCondensMatter,Roth2010_JPhysCondensMatter,Tarazona2008_incollection,Wu2007_AnnuRevPhysChem}. 
In a one-component system, though classical, these particles are indistinguishable in principle according to standard statistical mechanics \cite{Hansen2006_book}, which implies that the interaction between two particles is the same for any pair of particles provided they are at the same separation. 
This standard assumption breaks down for particles embedded in an elastic polymeric gel, if the particles are anchored to the surrounding gel matrix and/or cannot diffuse or propagate through it. 
In this case, the particles can be labeled according to their position in the matrix with their interaction energy persistently depending on the labeling. 
Thus, they are distinguishable. 
As a basic example, this situation is encountered for a simple bead-spring model, where the springs represent the elasticity and connectivity provided by the matrix and the beads represent the particles. 

Particle distinguishability leads to a different combinatorial prefactor in the classical partition function and therefore affects the entropy \cite{Gibbs1957_book}. 
However, at high density, a fluid of indistinguishable particles typically undergoes a freezing transition into a crystal. 
At low temperature, this crystal can be modeled by a  harmonic solid \cite{Chaikin2000_book,Espanol1996_PhysRevE,Jancovici1967_PhysRevLett}, where the neighboring particles are connected by springs.
In fact, this effective model of distinguishable particles provides a good approximation as the free energy of the system is dominated by the particle interactions that overwhelm the combinatorial contribution.

In the present paper, we exploit this idea to introduce a density functional approach for ferrogels and related systems.
Such ferrogels consist of magnetic colloidal particles that are embedded in a polymeric gel matrix \cite{filipcsei2007magnetic,ilg2013stimuli,menzel2015tuned, odenbach2016microstructure,lopez2016mechanics}. 
Examples of similar materials are given by magnetic elastomers, magnetorheological elastomers, or magnetosensitive elastomers, where often these terms are used interchangeably. 
Remarkably, the structural properties and rheological behavior of these materials are governed by a competition between the magnetic particle--particle interactions and the mechanical elasticity of the embedding polymeric gel matrix. 
As a consequence, it is possible to tune their properties during application by modifying the magnetic interactions via external magnetic fields. 
Therefore, these magnetorheological systems have many prospective and promising applications, such as tunable dampers \cite{sun2008study} or vibration absorbers \cite{deng2006development}. 

The theoretical description of these materials is challenging. 
While the specific properties arise on the mesoscopic colloidal particle scale, for practical applications one is interested in the overall macroscopic response.
To connect these scales in simulations, large numbers of individual particles need to be covered. 
For this purpose, recent work has focused on simplified minimal models. 
Starting on the microscale, at most a few individual polymer chains are resolved by coarse-grained bead-spring models \cite{weeber2012deformation,ryzhkov2015coarse,weeber2015ferrogels}. 
In still more reduced mesoscopic dipole-spring models, the elasticity of the matrix is directly represented by effective spring-like interactions between the particles, combined with long-ranged magnetic dipolar interactions between them \cite{annunziata2013hardening,pessot2014structural,tarama2014tunable,ivaneyko2015dynamic,pessot2016dynamic}. 
More explicit approaches treat the matrix directly by continuum elasticity theory, yet at the price of reduced accessible overall particle numbers \cite{han2013field,cremer2015tailoring,cremer2016superelastic,metsch2016numerical}. A kind of compromise between the two concepts can be found in Refs.~\onlinecite{biller2014modeling} and \onlinecite{biller2015mesoscopic}. 
Previous analytical approaches to link the different scales often relied on substantially simplifying idealizations concerning the positional particle configurations \cite{ivaneyko2011magneto,menzel2014bridging,zubarev2016towards}. 
Therefore, it is desirable to develop statistical means that allow for a more profound connection between the different scales in the future. 
As a step in this direction, we now suggest to employ the framework of classical density functional theory for a characterization of these complex materials.

Here we mainly follow the dipole-spring concept of distinguishable particles often used for the description of ferrogels. 
In order to keep the models simple, we study effective one-dimensional set-ups. 
Such a situation is realized, for instance, for elongated magnetic particle chains embedded into an elastic matrix \cite{huang2016buckling}, but also for magnetic filaments \cite{sanchez2013effects} made, \textit{e.g.}, of magnetic colloidal particles connected by DNA polymer strands \cite{dreyfus2005microscopic}. 
We map this system with its particle-distinguishing connectivity onto another one with an effective connectivity and indistinguishable particles \cite{Denton2007_inbook}.
Based on the considerations above, one expects a good agreement between real and effective connectivity at least for strong particle--particle interactions.
We use Monte-Carlo computer simulations of both situations and confirm that the results agree at high packing fractions and/or strong particle interactions. 
This opens the way to employ statistical-mechanical theories like classical density functional theory to also describe systems of particles that are, in principle, distinguishable. 
For the one-dimensional model, we use the exact Percus free-energy functional for hard rods \cite{Percus1976_JStatPhys} combined with a mean-field theory for the elastic and dipolar interactions and minimize the resulting grand canonical free energy functional with respect to the equilibrium one-body density field. 

We study two different models. 
In the first one, the elastic matrix is represented by harmonic springs between nearest-neighboring particles. 
Including thermal fluctuations, such a simple one-dimensional bead-spring model cannot show a phase transition \cite{VanHove1952_Physica,Cuesta2004_JStatPhys}. 
Thermal fluctuations have a strong impact in one spatial dimension and fuel the Landau-Peierls instability \cite{Landau2008_UkrJPhys,Peierls1934_HelvPhysActa,Chaikin2000_book}, which impedes periodic ordering. 
This fact is captured by our Monte-Carlo simulations, which take all contributions by thermal fluctuations into account exactly.  
Our mean-field density functional theory, however, introduces an artificial crystallization at low temperatures. 
Still, at higher temperatures we can obtain qualitative agreement between density functional theory and Monte-Carlo simulations for the pressure and compressibility of the system. 
These provide key material properties for the practical use of ferrogels. 

Later in this paper, we turn to an extended model, including an additional external elastic pinning potential for the colloidal magnetic particles. 
Such pinning potentials arise when the particles are embedded in a three-dimensional elastic bulk matrix \cite{huang2016buckling}. 
Similarly, the displacement of one embedded particle results in a matrix-mediated force on all other particles, so that we have additional long-ranged elastic particle--particle interactions.
In combination with the pinning potential, this suppresses the Landau-Peierls instability even though our model is effectively one-dimensional. 
Consequently, our density functional theory immediately shows much better agreement with Monte-Carlo simulations for both, the model with real connectivity and the version mapped towards indistinguishable particles.
During the synthesis of ferrogels such permanent straight one-dimensional magnetic particle chains embedded within three-dimensional ferrogel blocks are readily generated by applying external magnetic fields during the manufacturing process \cite{collin2003frozen,varga2003smart,gunther2012xray,borbath2012xmuct,gundermann2013comparison,huang2016buckling,Gundermann2017_SmartMaterStruct}. 

We remark at this stage that the problem of mapping from distinguishable to indistinguishable particles also occurs in density functional descriptions of polymeric bead models \cite{Chandler1986_JChemPhys_1,Chandler1986_JChemPhys_2}.
Typically, in tangential bead models for hard spheres \cite{Gu2003_JChemPhys,Slyk2016_JPhysCondensMatter}, one neglects the linking constraints of the chain and maps the excess free energy of the system onto an unconstrained hard-sphere fluid.

Our analysis paves the way to a future application of density functional theory of freezing also to two- and three-dimensional ferrogel models.
There, we anticipate thermal fluctuation effects to be of less influence, leading to a better agreement with simulations. 
It will further be useful to characterize other particulate systems embedded in a permanent elastic  matrix such as electrorheological elastomers \cite{an2003actuating,allahyarov2015simulation} or possibly even drug carriers and compartments within biological tissue \cite{tietze2013efficient}.

This paper is organized as follows. 
In Sec.~\ref{Sec.Dipole-spring_model}, we describe our first one-dimensional dipole-spring model and offer a method to map the real connectivity to an effective one.
Next, we describe the various methods used to study our systems in Sec.~\ref{Sec.Methods}. 
These methods are mainly mean-field density functional theory and canonical Monte-Carlo simulations. 
The supplemental material \cite{supplemental} contains an additional treatment using the Zerah-Hansen liquid-integral equation. 
Then we complete the discussion of the first dipole-spring model in Sec.~\ref{Sec.Results}, where we present results from our density functional theory and compare them to Monte-Carlo simulations. 
Subsequently, we proceed to our extended model in Sec.~\ref{Sec.Embedding_into_the_elastic_matrix}. 
Following a motivation of this extended model, we again compare results from density functional theory and Monte-Carlo simulations, showing their improved agreement. 
Finally, in Sec.~\ref{Sec.Conclusions}, we revisit our overall approach and discuss prospective uses and extensions beyond the one-dimensional models discussed here. 

\section{Dipole-spring model}
\label{Sec.Dipole-spring_model}

\begin{figure}
	\includegraphics[width=1.0\columnwidth]{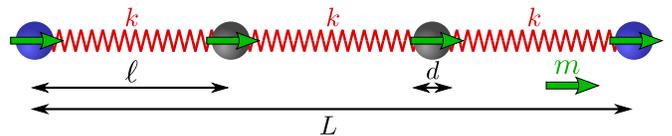}
	\caption{Sketch of our one-dimensional dipole-spring model for a ferrogel. Two outer particles (blue) form the system boundary and are at a fixed distance $L$. Additionally, there are $N$ mobile particles (dark gray) in between. Each particle carries a magnetic dipole moment of magnitude $m$, all of which point into the same direction along the system axis. Finally, harmonic springs of spring constant $k$ and equilibrium length $\ell$ connect each particle to its nearest neighbors.}
	\label{Fig.dipole_spring_model}
\end{figure}

We consider the following one-dimensional dipole-spring model, which is sketched in Fig.~\ref{Fig.dipole_spring_model}.
There are two outer particles at a fixed distance $L$, forming the system boundary, and $N$ mobile particles in between. 
All particles have a hard core of diameter $d$, which limits the closest approach of two particle centers to this distance. 
Additionally, all particles carry magnetic dipole moments of magnitude $m$ that all point in the same direction aligned with the system axis. 
Finally, each particle is connected to its nearest neighbors by a harmonic spring of spring constant $k$ and equilibrium length $\ell$.

As we will discuss below, the connectivity introduced by the harmonic springs renders the particles distinguishable.
We label the particles with indices $i = 0, \dots, N+1$ according to their position $x_i$ in ascending order.
The indices $i = 0$ and $i = N+1$ are used for the left and right boundary particles, respectively.
The total potential energy of the system consists of three contributions
\begin{equation}
	U = U_\textrm{h} + U_\textrm{m} + U_\textrm{e}, 
	\label{Eq.dipole_spring_model_potential}
\end{equation}
\textit{i.e.}, the hard core repulsion $U_\textrm{h}$, the magnetic dipolar interaction $U_\textrm{m}$, and the elastic interaction $U_\textrm{e}$.
We can write the former two as sums over the interactions between all particle pairs $i,j$ with $j > i$ 
\begin{equation}
	 U_\textrm{h} = \sum_{i = 0}^{N+1} \sum_{j > i} u_\textrm{h}(x_{ij}) \, ; \quad  u_\textrm{h}(x) = \begin{cases}
		\infty	& \textnormal{for} \,\,	x < d \, ,	\\
		0	& \textnormal{for} \,\,	x \geq d \, , 
	\end{cases}
	\label{Eq.hard_interaction}
\end{equation}
\begin{equation}
	U_\textrm{m} = \sum_{i = 0}^{N+1} \sum_{j > i} u_\textrm{m}(x_{ij}) \, ; \quad u_\textrm{m}(x) =  -2 \frac{\mu_0}{4\pi} \frac{m^2}{x^3} \, ,
	\label{Eq.magnetic_interaction}
\end{equation}
where $\mu_0$ is the vacuum permeability and $x_{ij} := |x_j - x_i|$ is the distance between a pair of particles. 
Since the pair interactions $u_\textrm{h}(x)$ and $u_\textrm{m}(x)$ do not depend on any particular labeling of the particles, the total interactions $U_\textrm{h}$ and $U_\textrm{m}$ are invariant under a relabeling of all particles. 
In contrast to that, the elastic interactions between nearest-neighbors
\begin{equation}
	U_\textrm{e} = \frac{k}{2} \sum_{i = 0}^N (x_{i,i+1} - \ell)^2 
	\label{Eq.elastic_interaction}
\end{equation} 
persistently depend on the labeling and therefore render the particles distinguishable.

%

\begin{figure}
	\includegraphics[width=1.0\columnwidth]{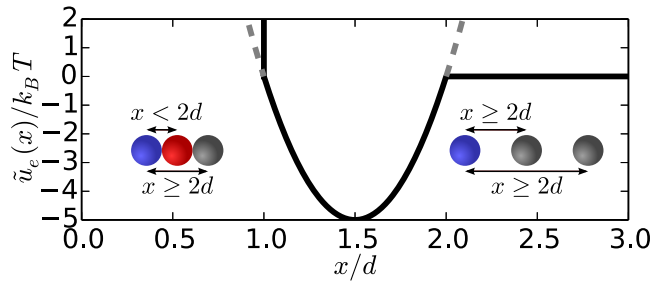}
	\caption{Illustration of how the pseudo-spring pair potential $\tilde{u}_\textrm{e}(x)$ combined with the hard-core repulsion $u_\textrm{h}(x)$ serves to approximate the effect of harmonic springs between nearest neighbors. A harmonic well of spring constant $k$ (here $k = 40 k_BT / d^2$) is centered around a distance $\ell = 1.5d$. It is cut at a distance $x = 2d$ and shifted to zero in order to confine the interaction to nearest neighbors only. The sketch on the left depicts a situation where a particle (blue) interacts with its nearest neighbor (red), as both particles are at a distance $x < 2d$. In the sketch on the right no interaction takes place since the distance is $x \geq 2d$. In any case, only nearest neighbors can ever interact as we always have $x \geq 2d$ for all other pairs of particles.}
	\label{Fig.Vpair}
\end{figure}

To facilitate a description of these systems with the tools of statistical mechanics, we map it onto a system of indistinguishable particles. 
This can be achieved by replacing the elastic interaction \eqref{Eq.elastic_interaction} with an approximative potential $\tilde{U}_\textrm{e}$ that can be decomposed into pairwise interactions $\tilde{u}_\textrm{e}(x)$. 
Ideally, such an approximative potential should still affect only nearest neighbors and provide the same result as Eq.~\eqref{Eq.elastic_interaction} under realistic circumstances.
We make the choice
\begin{equation}
	\begin{gathered}
		\tilde{U}_\textrm{e} = \sum_{i = 0}^{N+1} \sum_{j > i} \tilde{u}_\textrm{e}(x_{ij}) \, ; \\
		\tilde{u}_\textrm{e}(x) = \begin{cases}
			\frac{k}{2} \left[ (x - \ell)^2 - (2d - \ell)^2 \right]	&	\textrm{for} \,\, x < 2d \, , \\
			0	&	\textnormal{for} \,\, x  \geq 2d \, .
		\end{cases}
	\end{gathered}
	\label{Eq.pseudoelastic_interaction}
\end{equation}
The ``pseudo-spring'' pair potential $\tilde{u}_\textrm{e}(x)$ is illustrated in Fig.~\ref{Fig.Vpair} and consists of a harmonic well of spring constant $k$ centered around a distance $\ell < 2d$. 
The harmonic well is cut and shifted to zero potential strength at a distance $x \geq 2d$. 
For two particles at a distance $x < 2d$, this potential acts as a common harmonic spring.
Beyond this distance, the spring ``breaks'' leading to zero interaction.
The combination with the hard-core repulsion $u_\textrm{h}(x)$ in Eq.~\eqref{Eq.hard_interaction} limits the possible harmonic interaction to pairs of nearest-neighbors. 
Only nearest neighbors can be at a distance $x < 2d$. 
Next-nearest neighbors are always at a greater distance and, thus, excluded from the interaction. 

%

In the following, we refer to this potential as ``pseudo-spring'' interaction as opposed to the ``real-spring'' permanent connectivity between nearest neighbors.
Good agreement between the real-spring system and its mapped version using pseudo-springs can be expected in situations where the pseudo-springs do not break.
First, this is  the case at high packing fraction, when the confinement enforces small distances between nearest neighbors.
The packing fraction in our finite system is defined as
\begin{equation}
	\phi = \frac{N d}{L - d} \, ,
	\label{Eq.packing_fraction}
\end{equation}
because $L - d$ is the system length enclosed between the two hard boundary particles. 
At $\phi > (L-2d) / (L-d)$, the distance between nearest neighbors is smaller than $2d$ everywhere, such that breaking of pseudo-springs becomes impossible. 
Another limit is reached at high potential strength (large value of $k$) and moderate packing fraction of $\phi \gtrsim d / \ell$. 
Under these conditions, the harmonic well, as illustrated in Fig.~\ref{Fig.Vpair}, is deep compared to the thermal energy $k_BT$ and the system is sufficiently filled such that all particles are effectively trapped in the harmonic wells created by their nearest neighbors. 

For fixed values of $L$, $N$, and $d$, the physical input parameters determining all interactions are the magnetic moment $m$, the spring constant $k$, and the spring equilibrium length $\ell$. 
From now on, we measure all energies in units of $k_BT$ and all lengths in units of the particle diameter $d$. 
This implies to measure the spring constant in units of $k_0 = k_BT / d^2$ and the magnetic moment in units of $m_0 = \sqrt{\frac{4\pi}{\mu_0} k_BT d^3}$, while the
pressure and the compression modulus are given in units of $p_0 = K_0 = k_BT / d$.

\section{Methods}
\label{Sec.Methods}

We use three different methods to study our dipole-spring model. 
The first and most notable one is our density functional theory (DFT) description, for which we use the pseudo-spring approximation to make particles indistinguishable. 
Second, we perform canonical Monte-Carlo (MC) simulations for real springs as well as for pseudo-springs as a benchmark to test our DFT results. 
Finally, we have also solved the Zerah-Hansen liquid-integral equation to show that our pseudo-spring approximation is meaningful beyond the scope of DFT, see the supplemental material \cite{supplemental} for results and a description of the method.

\subsection{Density functional theory}
The central statement of classical DFT is that for a fixed temperature $T$ and interparticle pair potential $u(x)$, the Helmholtz free energy $\mathcal{F}[\rho]$ is a unique functional of the one-body density distribution $\rho(x)$. 
Likewise, there is a unique grand canonical free energy functional $\Omega[\rho]$ describing the system when it is exposed to an external potential $U_\textrm{ext}(x)$ and a particle reservoir at chemical potential $\mu$.
This grand canonical free energy functional has the form \cite{Tarazona2008_incollection}
\begin{equation}
	\Omega[\rho] = \mathcal{F}[\rho] + \int_0^L \rho(x) \big( U_\textrm{ext}(x) - \mu \big) \, dx
	\label{Eq.DFT_grand_functional}
\end{equation}
and is minimized by the equilibrium one-body density profile $\rho_\textrm{eq}(x)$. 
The minimum $\Omega[\rho_\textrm{eq}]$ corresponds to the thermodynamic grand canonical free energy in equilibrium.

Unfortunately, the exact free energy functional $\mathcal{F}[\rho]$ is usually unknown, so that one has to resort to approximations. 
These approximations usually start by splitting the free energy functional $\mathcal{F}[\rho] = \mathcal{F}_\textrm{id}[\rho] + \mathcal{F}_\textrm{ex}[\rho]$ into the exact free energy for the ideal gas
\begin{equation}
	\mathcal{F}_\textrm{id}[\rho] = k_BT \int_0^L \rho(x) \bigg( \ln\big( \Lambda \rho(x) \big) - 1 \bigg) \, dx \,
	\label{Eq.DFT_ideal_functional}
\end{equation}
with $\Lambda$ the thermal de Broglie wavelength, plus an excess contribution $\mathcal{F}_\textrm{ex}[\rho]$. 
For some special problems in one spatial dimension, the exact excess contribution can be derived \cite{Tutschka2000_PhysRevE}.
One such example is the Percus excess functional \cite{Percus1976_JStatPhys} for the one-dimensional hard-rod fluid,
\begin{equation}
	\begin{gathered}
		\begin{aligned}
			\mathcal{F}_\textrm{ex}^\textrm{P}[\rho] = -k_BT \int_{0}^L & \frac{\rho(x + d/2) + \rho(x - d/2)}{2} \\
			&\quad\times\, \ln\big( 1 - \eta(x) \big) \, dx \, , &
		\end{aligned} \\
		\textrm{with } \eta(x) = \int_{x-d/2}^{x+d/2} \rho(x') \, dx' \, .
	\end{gathered}
	\label{Eq.DFT_Percus_functional}
\end{equation} 
It takes one-dimensional hard repulsions exactly into account and, thus, provides a good starting point for the construction of a functional describing our dipole-spring model.
Here, we combine it with an approximate mean-field excess functional accounting for the soft pair interactions consisting of our pseudo-spring pair potential $\tilde{u}_\textrm{e}(x)$ and the magnetic dipolar pair interaction $u_\textrm{m}(x)$,
\begin{equation}
		\begin{aligned}
			\mathcal{F}_\textrm{ex}^\textrm{MF}[\rho] &= \int_0^L  \int_0^L \, \big( \tilde{u}_\textrm{e}(|x - x'|) + u_\textrm{m}(|x - x'|) \big) \\
			&\qquad \times g(|x - x'|) \rho(x) \rho(x') \, dx' \, dx \, ,
		\end{aligned}
	\label{Eq.DFT_meanfield_functional}
\end{equation}
where the distribution function $g(x)$ satisfies the no-overlap condition $g(x) = 0$ for $x < d$. 
The mean-field approximation assumes that the pair potentials are soft enough to regard the particle positions as basically uncorrelated \cite{Tarazona2008_incollection}.
Here we make the simplifying assumption that $g(x) = 1$ for all distances $x > d$.
In total, our free energy functional is given by
\begin{equation}
	\mathcal{F}[\rho] = \mathcal{F}_\textrm{id}[\rho] + \mathcal{F}_\textrm{ex}^\textrm{P}[\rho] + \mathcal{F}_\textrm{ex}^\textrm{MF}[\rho] \, . 
	\label{Eq.DFT_dipole-spring_functional}
\end{equation}

The boundary of our finite systems consists of the leftmost and rightmost particles, which are fixed but otherwise identical to the enclosed particles, see Fig.~\ref{Fig.dipole_spring_model}.
Their influence on the enclosed density profile enters via an external potential
\begin{equation}
	U_\textrm{ext}(x) = u(L - x) + u(x) \, ,
	\label{Eq.DFT_external_potential}
\end{equation}
where $u(x) = u_\textrm{h}(x) + u_\textrm{m}(x) + \tilde{u}_\textrm{e}(x)$.
This completes our grand canonical free energy functional $\Omega[\rho]$. 

Functional derivation of Eq.~\eqref{Eq.DFT_grand_functional} leads to the Euler-Lagrange equation
\begin{equation}
	\frac{\delta \Omega[\rho]}{\delta \rho(x)} = \frac{\delta \mathcal{F}[\rho]}{\delta \rho(x)} + U_\textrm{ext}(x) - \mu \overset{!}{=} 0 \, , 
	\label{Eq.DFT_Euler-Lagrange}
\end{equation}
which can be used to determine the equilibrium density profile minimizing $\Omega[\rho]$.
In practice, however, we numerically calculate our equilibrium density profile $\rho(x)$ by performing a dynamical relaxation of $\Omega[\rho]$ \cite{Loewen1993_JChemPhys}. 
This scheme fixes the average particle number $\langle N \rangle$ instead of the chemical potential $\mu$ and is described in detail in the appendix. 

After the relaxation, we have access to the grand canonical free energy $\Omega$. 
This enables us to calculate a pressure $p = -\frac{\partial \Omega}{\partial L}\big\vert_{\langle N \rangle, T}$ and a compression modulus $K = -L \frac{\partial p}{\partial L}\big\vert_{\langle N \rangle, T}$
by varying the system length $L$ at fixed average particle number $\langle N \rangle$ and probing the corresponding change in $\Omega$.

\subsection{Monte-Carlo simulation}
We perform canonical Monte-Carlo (MC) simulations at fixed particle number $N$, system length $L$, and temperature $T$ \cite{Frenkel2001_book}.
After equilibrating the systems, we sample the pressure $p$, the compression modulus $K$, and the equilibrium density profile $\rho(x)$.
To sample the pressure, we affinely deform the system by a factor $(L + \Delta L)/L$ and probe the corresponding change in Helmholtz free energy $F(N,L,T)$. 
$\Delta L$ is a small change in system length.
It can be shown that the pressure is related to the acceptance ratio of such volume moves by \cite{Harismiadis1996_JChemPhys}
\begin{equation}
	\begin{aligned}
		p	&= -\frac{\partial F(N,L,T)}{\partial L} \approx -\frac{F(N,L + \Delta L,T) - F(N,L,T)}{\Delta L} \\ 
			&=  \frac{k_BT}{\Delta L} \ln\left\langle \left(\frac{L + \Delta L}{L}\right)^N \exp(-\Delta U / k_BT) \right\rangle \, .
	\end{aligned}
	\label{Eq.Widom_pressure}
\end{equation}
$\Delta U$ is the change in system energy associated with the volume move and $\langle \cdot \rangle$ denotes the ensemble average. 
In order to capture the pressure contributions of the hard repulsions in our systems, the volume moves must be compressive ($\Delta L < 0$).
Given the pressure, the compression modulus can be calculated using $K = -L \frac{\partial p}{\partial L}\big\vert_{N,T}$. 

\section{Results}
\label{Sec.Results}

In the following, we present results for our one-dimensional dipole-spring model.
First we concentrate on a non-magnetic system to test the feasibility of the mapping onto indistinguishable particles. 
Then we add the magnetic interaction and discuss how this affects the density profile and the pressure in our systems.
Finally, we turn to the thermodynamic compression modulus, which is a key quantity to characterize ferrogel systems as it can be controlled by changing the magnetic properties.

\subsection{Non-magnetic system}
\begin{figure}[b]
	\includegraphics[width=1.0\columnwidth]{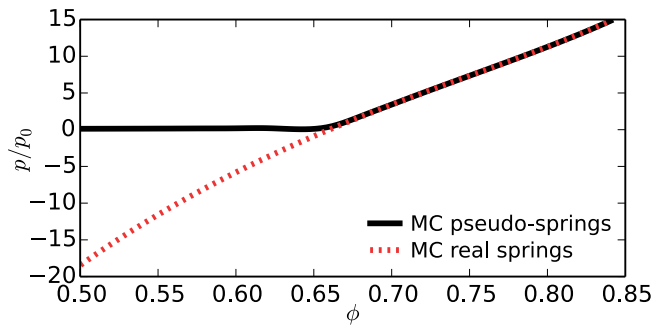}
	\caption{Comparison of the equations of state from MC-simulations of the real-spring system with the mapped version using pseudo-springs instead. In both systems we have $L = 51 d$, $k = 40 k_0$ and $\ell = 1.5d$. At a packing fraction $\phi \gtrsim d / \ell = \frac{2}{3}$, the pseudo-spring system is filled with particles trapped in the harmonic wells of their nearest neighbors and, thus, behaves essentially identical to the system featuring real springs. For lower packing fractions in the real-spring system, the springs between nearest neighbors are stretched on average, so that the system would contract if the boundaries were not fixed. Thus, the pressure is negative for these packing fractions.}
	\label{Fig.real_vs_pseudo_springs}
\end{figure}

First of all, we confirm within our MC-simulations that the pseudo-spring pair potential is an appropriate replacement for real springs between nearest neighbors. 
Figure~\ref{Fig.real_vs_pseudo_springs} compares the equations of state $p(\phi)$ for both situations in a system of length $L = 51 d$. 
The spring parameters $k = 40 k_0$ and $\ell = 1.5d$ are the same as in Fig.~\ref{Fig.Vpair}. 
Using these parameters, we can confirm that at packing fractions $\phi \gtrsim d / \ell = \frac{2}{3}$ the mapping to indistinguishable particles using pseudo-springs works well.

\begin{figure}
	\includegraphics[width=1.0\columnwidth]{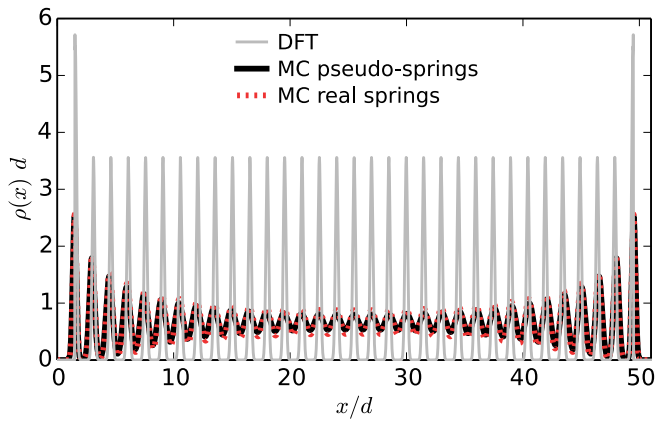}
	\caption{Density profiles $\rho(x)$ obtained from real-spring and pseudo-spring MC as well as from DFT calculations at a packing fraction $\phi = \frac{2}{3}$ and otherwise with the same parameters as in Fig.~\ref{Fig.real_vs_pseudo_springs}. For this packing fraction, the two MC-simulations are in good agreement and show a liquid-like behavior as expected in one spatial dimension. However, the density profile obtained from DFT is qualitatively different and displays an artificial crystalline behavior.}
	\label{Fig.MC_vs_DFT_densityprofile_k40}
\end{figure}

Let us now compare MC and DFT results using the same parameters.
Figure~\ref{Fig.MC_vs_DFT_densityprofile_k40} shows three density profiles $\rho(x)$ at a packing fraction $\phi = \frac{2}{3}$, one from the real-spring MC, one from the pseudo-spring MC, and one from DFT.
While the two MC density profiles expectedly agree with each other and display a liquid-like behavior near rigid boundaries, the DFT density profile is qualitatively different and resembles a crystal.

\begin{figure}
	\includegraphics[width=1.0\columnwidth]{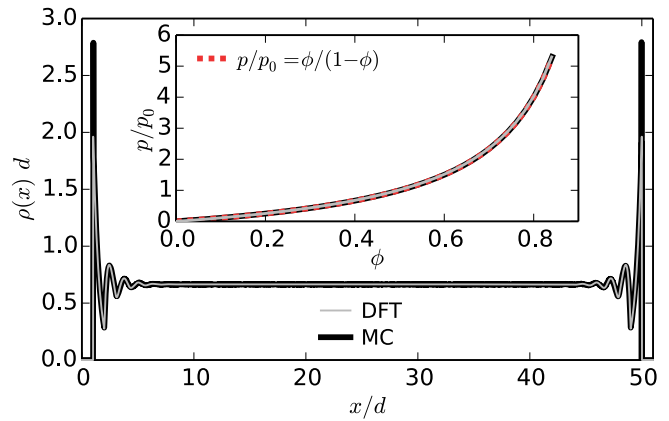}
	\caption{When setting $k = 0$ and $m = 0$, we recover the hard-rod fluid and observe perfect agreement between MC and DFT. In this case, both density profiles $\rho(x)$ show liquid-like behavior and the equations of state (inset) match the exact result $p(\phi) = \frac{\phi}{1 - \phi} p_0$ \cite{Percus1976_JStatPhys}. This demonstrates that the mean-field term in Eq.~\eqref{Eq.DFT_meanfield_functional} is responsible for the disagreement between DFT and MC, as it disregards some of the contributions by thermal fluctuations.}
	\label{Fig.MC_vs_DFT_hardrods}
\end{figure}

This crystalline appearance displayed by the DFT is unphysical.
Our system is one-dimensional, all particle interactions are short-ranged and there are no external fields. 
For such systems, the existence of a phase transition can be ruled out \cite{Cuesta2004_JStatPhys,Dyson1969_CummunMathPhys,Ruelle1968_CommunMathPhys,VanHove1952_Physica}.
This fact is accurately captured by our MC-simulations that display a liquid phase even for this high value of $k = 40 k_0$, as they explicitly include all effects of thermal fluctuations.
In one spatial dimension, thermal fluctuations have a particularly strong effect. 
They can escalate into long-ranged fluctuations scaling in amplitude with the system size, capable of destroying periodic ordering. 
This is the well-known Landau-Peierls instability \cite{Landau2008_UkrJPhys,Peierls1934_HelvPhysActa,Chaikin2000_book}. 
Within the DFT, some thermal fluctuations are introduced by the ideal gas term [see Eq.~\eqref{Eq.DFT_ideal_functional}], which pushes the system towards disorder. 
However, the mean-field term [see Eq.~\eqref{Eq.DFT_meanfield_functional}] excludes other contributions by fluctuations.
We conclude that this term is responsible for the unphysical crystallization. 
Our conjecture is supported by setting $k = 0$ and $m = 0$, \textit{i.e.} setting the mean-field term to zero.
Then, we recover the hard-rod fluid also for the DFT and find perfect agreement with MC-simulations, see Fig.~\ref{Fig.MC_vs_DFT_hardrods}.

The Landau-Peierls instability is well-known to be most prominent in one spatial dimension. 
In future studies in two and three dimensions, we therefore expect a significantly weaker effect of the thermal fluctuations, which should lead to a better agreement between simulations and mean-field DFT.  
For now, we achieve qualitative agreement between DFT and MC by raising the temperature (which means decreasing $k$) until the DFT system enters the liquid state. 

\begin{figure}
	\includegraphics[width=1.0\columnwidth]{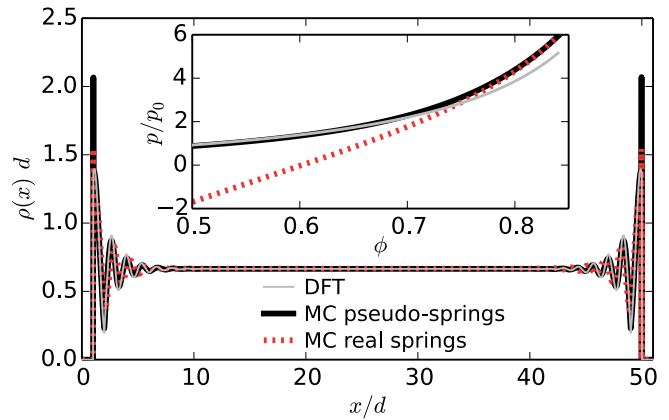}
	\caption{Density profiles as in Fig.~\ref{Fig.MC_vs_DFT_densityprofile_k40} but using  a ten times lower spring constant $k = 4 k_0$. Now the depth of the harmonic well of the pseudo-spring pair potential is of the order of $k_BT$ such that pseudo-springs frequently break. As a result, DFT and pseudo-spring MC both show liquid-like behavior and are in much better agreement. However, this comes at the price of worse agreement between the  pseudo-spring MC and the real-spring MC. The inset shows equations of state  $p(\phi)$ for these three systems which confirm these observations. There is agreement between DFT and the pseudo-spring MC at least in the range around $\phi \approx \frac{2}{3}$ but the pseudo-spring and real-spring MC only agree at very high packing fractions.}
	\label{Fig.MC_vs_DFT_k4_mm0}
\end{figure}

Figure~\ref{Fig.MC_vs_DFT_k4_mm0} shows a comparison between density profiles as well as equations of state for the same systems as in Fig.~\ref{Fig.MC_vs_DFT_densityprofile_k40}, but with a ten times lower spring constant $k = 4 k_0$. 
The depth of the harmonic well in the pseudo-spring potential is now of the order of $k_BT$ so that breaking of pseudo-springs is a common event. 
This renders the density profile in DFT more liquid-like, which improves the agreement with the pseudo-spring MC substantially.
At the same time though, the pseudo-spring mapping becomes a bad approximation for the real connectivity. 
Only at high packing fractions, where the confinement prevents pseudo-spring breaking, we can reach agreement between the pseudo-spring and real-spring MC again.


%
\begin{figure*}[t]
	\includegraphics[width=1.0\textwidth]{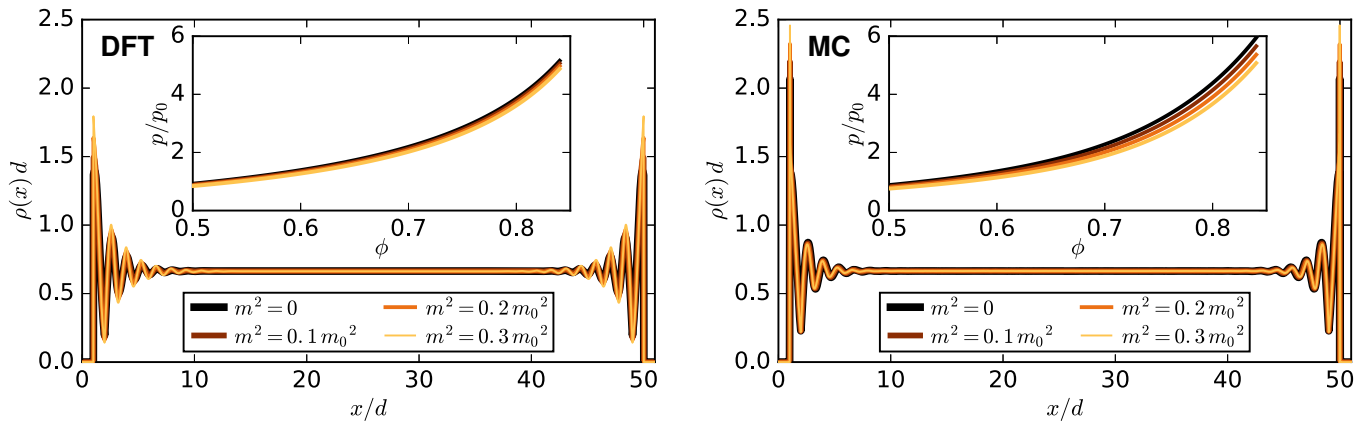}
	\caption{Density profile $\rho(x)$ as well as equation of state $p(\phi)$ obtained from DFT and pseudo-spring MC upon activating the magnetic moments. In the DFT, all density peaks increase in amplitude. On the contrary, in the MC only the first peak is affected. This is probably again due to the underestimation of thermal fluctuations in our mean-field DFT. The more patterned structure puts less emphasis on configurations where two particles are close and magnetic attractions are strong. Thus, the DFT predicts a much smaller downward shift for the pressure $p$ when increasing the magnetic moments than the MC (insets).}
	\label{Fig.MC_vs_DFT_k4_mm}
\end{figure*}

\subsection{Influence of magnetic interactions}

We now activate the magnetic dipolar interactions and discuss the resulting changes for our systems. 
Figure~\ref{Fig.MC_vs_DFT_k4_mm} demonstrates that increasing $m$ increases the amplitudes of all peaks in the DFT, whereas in the pseudo-spring MC only the first peak is affected.

Again, our mean-field DFT seems to overestimate the tendency to form a patterned structure because of its incomplete representation of thermal fluctuations. 
The reason is that, effectively, the particles do not fluctuate as much around their average positions and do not come as close to each other, where the pseudo-spring interaction and the dipolar interaction increase (the latter with inverse cubic distance).
As a consequence, the DFT underestimates the averaged strength of the pair interactions in the system. 
This becomes apparent in the equation of state, where the MC predicts a much stronger downwards shift when increasing the magnetic moments (see the insets of Fig.~\ref{Fig.MC_vs_DFT_k4_mm}). 

As a liquid-state approach alternative to DFT, we have also solved the Zerah-Hansen liquid-integral equation. 
Corresponding results in comparison to MC-simulations can be found in the supplemental material \cite{supplemental}.

\subsection{Thermodynamic compression moduli}
\begin{figure*}[t]
	\includegraphics[width=1.0\textwidth]{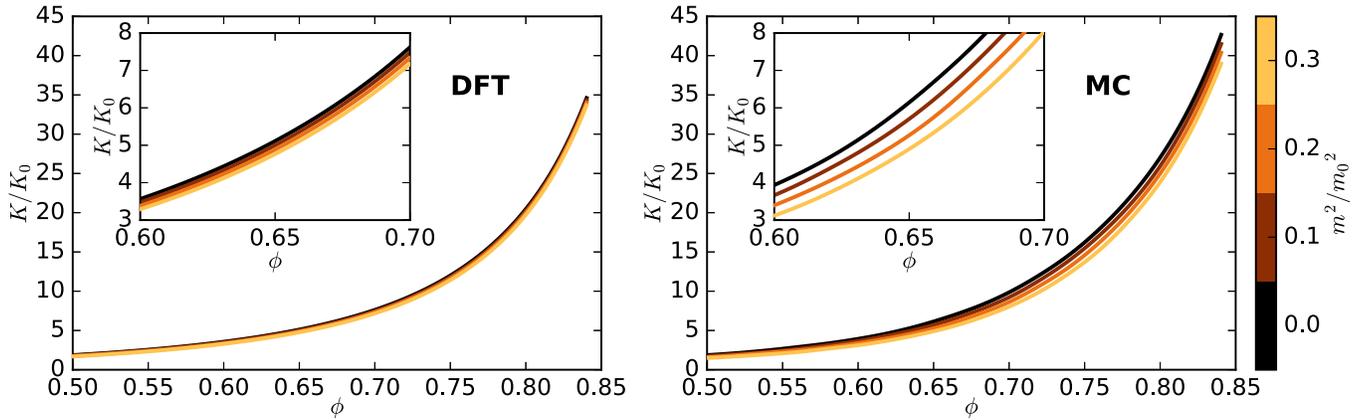}
	\caption{Compression modulus $K$ as a function of packing fraction $\phi$ for various magnetic moments. In the DFT calculations the curves are only very slightly shifted downwards when increasing the magnetic moment, a trend that we have already seen in the pressure in Fig.~\ref{Fig.MC_vs_DFT_k4_mm}. Furthermore, compared to the pseudo-spring MC-simulation, the compression modulus is overall lower, especially at high packing fractions. Again, this is due to the mean-field nature of our DFT, which underestimates the contributions of the magnetic and elastic pair interactions by not taking thermal fluctuations fully into account. In the MC-simulations, the particles deviate more in their positions, increasing the influence of these pair interactions on the compression modulus. Thus, increasing the magnetic moment has a greater effect.}
	\label{Fig.MC_vs_DFT_k4_compressionModulus_mm}
\end{figure*}

Finally, we evaluate the elastic moduli of the DFT and pseudo-spring MC systems for various magnitudes $m$ of the magnetic moment. 
We present them as a function of packing fraction $\phi$ in Fig.~\ref{Fig.MC_vs_DFT_k4_compressionModulus_mm}.
The DFT predicts only a very slight downward shift of the compression modulus when increasing the magnetic moment. 
In contrast to that, the shift is significantly more pronounced in the pseudo-spring MC. 
Additionally, the overall value of the compression modulus at high packing fractions is lower in the DFT.

These observations are in line with our earlier results. 
The mean-field DFT overestimates the tendency to form patterned structures. 
It therefore underestimates both, contributions by magnetic and elastic pair potentials. 
If the fluctuations of the particle positions were more pronounced, there would be more emphasis on configurations with strong elastic and magnetic interactions and their influence on the compression modulus would be stronger.

\section{Embedding into the elastic matrix}
\label{Sec.Embedding_into_the_elastic_matrix}
So far, we have considered a simple one-dimensional dipole-spring model. 
There, the elastic matrix is solely represented by springs between nearest-neighbor magnetic particles. 
Now we turn to an extended model, explicitly describing a single linear chain of magnetic particles that is embedded in a three-dimensional elastic matrix.

\subsection{Dipole-spring model for a linear embedded chain}

We begin by constructing an effective pinning potential $U_\textrm{mp}$ for the embedded particles within the three-dimensional matrix as well as an effective pair interaction $u_\textrm{pp}$ between two embedded particles \emph{mediated} by the matrix. 
Subsequently, we translate these potentials into a network of springs describing the overall elastic interactions.

If a single spherical particle of diameter $d$ embedded in an infinitely extended homogeneous elastic matrix is displaced by a vector $\Delta \mathbf{R}$, it distorts the elastic environment.
Then the restoring force $\mathbf{F}_\textrm{mp}$ that the matrix exerts onto the particle is given by \cite{Phan-Thien1993_JElasticity,Phan-Thien1994_ZAngewMathPhys,Puljiz2016_PhysRevLett,Puljiz2016_arXiv}
\begin{equation}
	\mathbf{F}_\textrm{mp}(\mathbf{U}) = - \frac{12\pi(1 - \nu) G d}{5 - 6\nu} \Delta\mathbf{R} \, ,
	\label{Eq.force_matrix-particle}
\end{equation}
where $\nu$ is the Poisson ratio, that equals $\nu=1/2$ for incompressible matrices, and $G$ is the shear modulus. 
The force can be connected via $\mathbf{F}_\textrm{mp} = -\nabla U_\textrm{mp}$ to a harmonic potential
\begin{equation}
	U_\textrm{mp}(\Delta\mathbf{R}) = \frac{1}{2} k_\textrm{mp} (\Delta\mathbf{R})^2 \, 
	\label{Eq.potential_matrix-particle}
\end{equation}
with the spring constant $k_\textrm{mp} := \big(12\pi(1 - \nu) G d \big) / \big(5 - 6\nu \big)$.

Now we consider two embedded particles, labeled as ``1'' and ``2'', respectively. 
Upon displacing these particles by vectors $\Delta \mathbf{R}_1$ and $\Delta \mathbf{R}_2$, they experience the forces $\mathbf{F}_1$ and $\mathbf{F}_2$. 
In our one-dimensional set-up, we only consider forces and displacements along the particle center-to-center vector $\mathbf{r}$. 
To first order in the particle distance, i.e.\, to order $1/r$ with $r=|\mathbf{r}|$ we then obtain \cite{Puljiz2016_PhysRevLett,Puljiz2016_arXiv,Phan-Thien1994_ZAngewMathPhys,Phan-Thien1993_JElasticity}: 
\begin{equation}
  \left( \begin{array}{c}\mathbf{F}_1\\[.1cm] \mathbf{F}_2\end{array} \right) =
  \left( \begin{array}{cc} -k_\textrm{mp} & \frac{k_\textrm{mp}^2}{4\pi G}\frac{1}{r}\\[.1cm]
  \frac{k_\textrm{mp}^2}{4\pi G}\frac{1}{r} & -k_\textrm{mp} \end{array} \right) \cdot 
  \left( \begin{array}{c}\Delta \mathbf{R}_1\\[.1cm] \Delta \mathbf{R}_2\end{array} \right) \, .
  \label{Eq.force-displacement-relation}
\end{equation}
Here, entries on the diagonal represent the restoring pinning forces \eqref{Eq.force_matrix-particle}. 
The off-diagonal contributions result from the matrix distortions that are caused by the displacement of one particle but affect the other embedded particle. 
To construct an effective pair potential, we here only consider symmetric situations where $\Delta \mathbf{R}_1 = -\Delta \mathbf{R}_2$.
Then, the change in distance between the two particles is $\Delta \mathbf{r} = \Delta \mathbf{R}_1 - \Delta \mathbf{R}_2 = 2\Delta \mathbf{R}_1$. 
Per particle, we can thus rewrite the effective matrix-mediated inter-particle interaction as a function of $\Delta \mathbf{r}$ in the form of an effective potential
\begin{equation}
	u_\textrm{pp}(\pm \Delta \mathbf{r}) = \frac{1}{2} k_\textrm{pp}(r)(\Delta \mathbf{r})^2 \, ,
	\label{Eq.matrix-mediated_pair-potential} 
\end{equation}
with a distance-dependent spring constant
\begin{equation}
	k_\textrm{pp}(r) := \frac{k_\textrm{mp}^2}{8\pi G}\frac{1}{r} = \frac{3}{2} \frac{1 - \nu}{5 - 6\nu} k_\textrm{mp} \frac{d}{r}. 
	\label{Eq.kpp_distance_dependent}
\end{equation}

\begin{figure}
	\includegraphics[width=1.0\columnwidth]{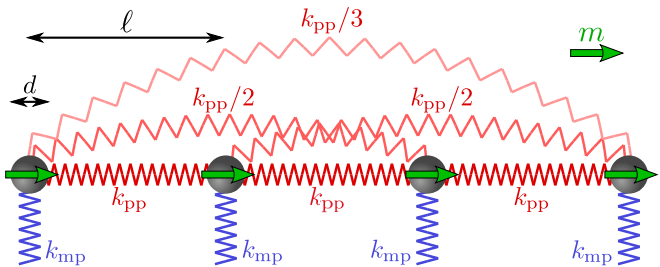}
	\caption{Sketch of our extended dipole-spring model for a one-dimensional chain of magnetic particles of diameter $d$ embedded into a three-dimensional elastic matrix. The elastic embedding is represented by a harmonic potential with spring constant $k_\textrm{mp}$, pinning each particle $i$ (labeled from left to right) to its initial position $x_i^0 := i \ell$. The elastic particle--particle interaction \emph{mediated} by the matrix is represented by the springs connecting the particles. Between nearest neighbors, there are springs of spring constant $k_\textrm{pp}$ and equilibrium length $\ell$. Next-nearest neighbors are connected by springs of spring constant $k_\textrm{pp}/2$ and equilibrium length $2\ell$, thereafter the parameters are $k_\textrm{pp}/3$ and $3\ell$, and so forth. Finally, all particles carry a fixed magnetic dipolar moment of magnitude $m$ aligned with the system axis.}
	\label{Fig.dipole_spring_model_embedded}
\end{figure}

Using the two potentials $U_\textrm{mp}$ and $u_\textrm{pp}$ as an input, we now motivate an extended dipole-spring model for a linear, initially homogeneous chain of $N$ particles embedded into an elastic matrix, see Fig.~\ref{Fig.dipole_spring_model_embedded}. 
We label the particles from left to right by $i = 1, \dots, N$, according to their equilibrium positions $x_i^0 := i \ell$ within the chain.
The total pinning potential based on Eq.~\eqref{Eq.potential_matrix-particle} then becomes 
\begin{equation}
	U_\textrm{mp} = \sum_{i = 1}^N \frac{1}{2} k_\textrm{mp} (x_i - i \ell)^2 \, .
	\label{Eq.potential_matrix-particle_total}
\end{equation}

To approximate the matrix-mediated particle--particle interactions between two particles $i,j$ we replace the $1/r$ dependence of the spring constant \eqref{Eq.kpp_distance_dependent} by $1 / |j-i| \ell$. 
Thus, we have for the total interaction between all pairs of particles
\begin{equation}
	U_\textrm{pp} = \sum_{i = 1}^N \sum_{j > i} \frac{1}{2} \frac{k_\textrm{pp}}{|j-i|} \big(x_{ij} - |j-i|\ell \big)^2\, , 
	\label{Eq.potential_particle-particle_total}
\end{equation}
where $k_\textrm{pp} := \frac{3}{2} \frac{1 - \nu}{5 - 6\nu} \frac{d}{\ell} k_\textrm{mp}$. 
Essentially, this means that each particle $i$ is connected to all other particles $j$ with harmonic springs of spring constant $k_\textrm{pp} / |j-i|$ and spring equilibrium length $|j - i|\ell$, see Fig.~\ref{Fig.dipole_spring_model_embedded}. 
From now on, we assume incompressibility of the elastic matrix and set $\nu = 1/2$. 

\begin{figure}
	\includegraphics[width=1.0\columnwidth]{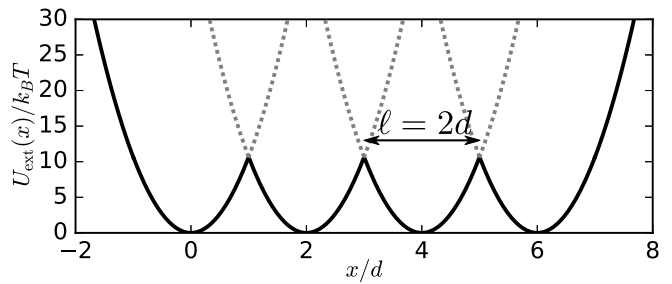}
	\caption{Illustration of the pseudo-spring external potential in Eq.~\eqref{Eq.embedded_Vext} representing the pinning effect of the embedding elastic matrix in our DFT. There are $N$ harmonic wells at a spacing $\ell$ with spring constant $k_\textrm{mp} = \frac{8\ell}{3d} k_\textrm{pp}$  (here $\ell = 2d$, $\nu = 1/2$, and  $k_\textrm{pp} = 4 k_0$, which translates to $G = \frac{8\ell}{9 \pi d^2} k_\textrm{pp} = \frac{64}{9\pi} \frac{k_0}{d}$). The corresponding harmonic potentials of the individual wells are cut where they overlap with the potential of another well. This leaves the leftmost and rightmost wells unbounded to the sides and, therefore, the whole particle chain remains confined.}
	\label{Fig.embedded_Vext}
\end{figure}

Of course, in this extended dipole-spring model the particles are again distinguishable by their positions. 
As before, it needs to be mapped to the use in our DFT. 
For this purpose, we replace the harmonic springs in the model by ``pseudo-springs'', following the ideas outlined in Sec.~\ref{Sec.Dipole-spring_model}. 
To include the pinning potential $U_\textrm{mp}$, we use an external potential consisting of a series of $N$ harmonic wells 
\begin{equation}
	U_\textrm{ext}(x) = \min_{i = 1, \dots, N} \left\lbrace \frac{1}{2} k_\textrm{mp} (x - i\ell)^2  \right\rbrace
	\label{Eq.embedded_Vext}
\end{equation}
as depicted in Fig.~\ref{Fig.embedded_Vext}.
\begin{figure}
	\includegraphics[width=1.0\columnwidth]{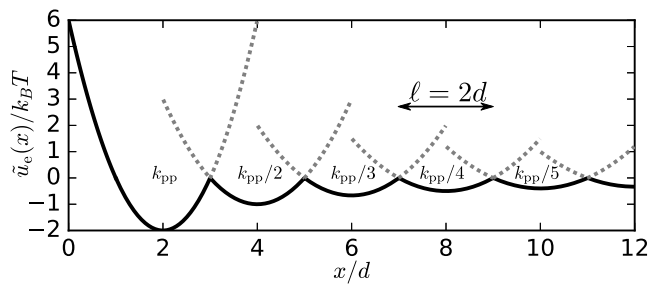}
	\caption{Effective elastic pair potential in Eq.~\eqref{Eq.embedded_Vpair} to represent the network of springs in Eq.~\eqref{Eq.potential_particle-particle_total} in our DFT calculations. The harmonic wells with width $\ell = 2d$ have a spring constant  $k_\textrm{pp}/i$, where $i$ is the index of the well and $k_\textrm{pp}= 4 k_0$.}
	\label{Fig.embedded_Vpair}
\end{figure}
To represent the network of springs in Eq.~\eqref{Eq.potential_particle-particle_total} between one particle and all other particles by ``pseudo-spring'' interactions, we use 
\begin{equation}
	\begin{gathered}
		\begin{aligned}
			\tilde{u}_\textrm{e}(x) = &\frac{k_\textrm{pp}}{2} \left[ (x - \ell)^2 - \frac{{\ell}^2}{4} \right] \mathds{1}_{[0, \frac{3}{2} \ell]}(x) \\
				+ \sum_{i=2}^\infty &\frac{k_\textrm{pp}}{2i} \left[ (x - i\ell)^2 - \frac{{\ell}^2}{4} \right] \mathds{1}_{[(i - \frac{1}{2})\ell, \, (i + \frac{1}{2})\ell]}(x) \, ,
		\end{aligned}
		\\
		\textrm{with } \mathds{1}_{[a,b]}(x) = \begin{cases}
			1	& \textnormal{for} \,\,	x \in [a,b] \, ,	\\
			0	& \textnormal{else} . 
		\end{cases}
	\end{gathered}
	\label{Eq.embedded_Vpair}
\end{equation}
This pair potential is illustrated in Fig.~\ref{Fig.embedded_Vpair} and consists of a series of harmonic wells.  
The spring constants of the wells decay with the neighbor number $i$ from the origin just like the individual springs connecting a particle to all other particles in the dipole-spring model, see Fig.~\ref{Fig.dipole_spring_model_embedded}.
Furthermore, the boundaries of the wells are shifted to zero so that we have a vanishing pair potential at infinite distance. 
Since the depth of the wells roughly decays as $1/x$, the interaction is long-ranged.

The width of the wells in both, the external and the pair potential, is given by $\ell$. 
This width should be larger than $d$ and here we choose $\ell = 2d$. 
Thus, both potentials in principle allow more than one particles to occupy a single well. 
However, the external potential pinning the particles to their equilibrium positions in the matrix is relatively strong. 
The particles should, therefore, remain centered in their respective wells on average. 

\subsection{Results}

\begin{figure*}[t]
	\includegraphics[width=1.0\textwidth]{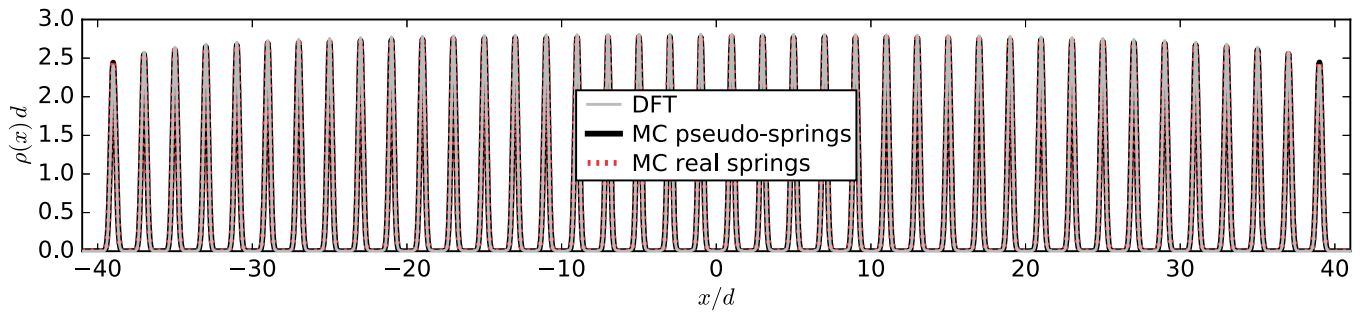}
	\caption{Density profiles for a non-magnetic chain of $N = 40$ particles at an equilibrium interparticle distance $\ell = 2d$ embedded in an elastic matrix ($k_\textrm{pp} = 4 k_0$, $k_\textrm{mp} =  \frac{16}{3} k_\textrm{pp}$, $m = 0$). The density profiles are sharply peaked around the equilibrium positions of the particles. There is good agreement between DFT, pseudo-spring MC and real-spring MC.}
	\label{Fig.MC_vs_DFT_embedded_densityprofile_mm0}
\end{figure*}
\begin{figure*}[t]
	\includegraphics[width=1.0\textwidth]{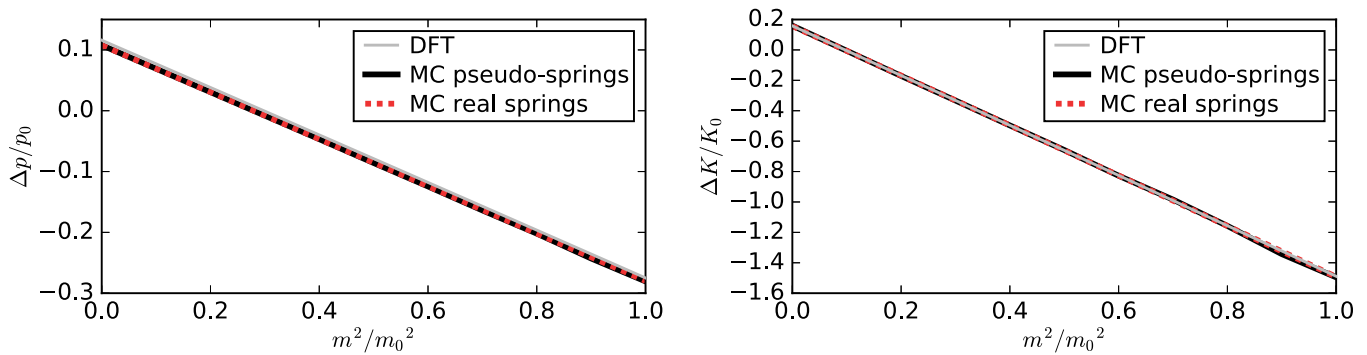}
	\caption{Pressure contribution $\Delta p$ and contribution $\Delta K$ to the compression modulus as a function of the squared magnetic dipole moment $m^2$. Both quantities show a linear monotonous decrease with $m^2$. The DFT shows a small offset in $\Delta p$ of the order of $0.01 p_0$ compared to the MC, but the slopes are almost identical. For the compression modulus, the deviations are of the order of $0.001 K_0$.}
	\label{Fig.MC_vs_DFT_embedded_p_and_K}
\end{figure*}

We now discuss our results for our extended dipole-spring model for a magnetic particle chain embedded in a three-dimensional elastic matrix.
To this end, we consider a chain of $N = 40$ particles, with an equilibrium interparticle distance $\ell = 2d$ and spring constants $k_\textrm{pp} = 4 k_0$, $k_\textrm{mp} = \frac{8\ell}{3d} k_\textrm{pp} = \frac{16}{3} k_\textrm{pp}$. 

To evaluate the contribution of the magnetic chain to the pressure and compression modulus as well as to evaluate the DFT numerically, we address one part of the elastic matrix of length $L = 100d$ that contains the magnetic chain. 
This choice of $L$ is arbitrary, the only requirement for $L$ is to be larger than the total equilibrium length $N\ell = 80d$ of the chain by a reasonable amount. 

Figure~\ref{Fig.MC_vs_DFT_embedded_densityprofile_mm0} shows density profiles obtained from DFT, pseudo-spring MC, and real-spring MC when setting the magnetic moment to $m = 0$. 
In contrast to our former dipole-spring model, periodic structures appear here in the density profiles resulting from all three methods, even though this model is still effectively one-dimensional. 
The reason is, first, that the elastic particle--particle interaction decays only slowly with the distance and is effectively long-ranged. 
This applies to both, the real-spring and the pseudo-spring version.
Second, we have a pinning potential suppressing large amplitude fluctuations of the particles around their pinning positions. 
Together, both contributions counteract the Landau-Peierls instability and can facilitate periodic structures also in one spatial dimension \cite{Cuesta2004_JStatPhys,Wong1997_Nature,Dyson1969_CummunMathPhys,Ruelle1968_CommunMathPhys,VanHove1952_Physica}. 
In this way, the role of thermal fluctuations is substantially reduced, and our mean-field DFT performs much better when compared to the MC-simulations. 

Let us now address the pressure and compression modulus. 
As before, they can  be determined by probing the energetic change of the system upon deformation. 
However, we keep in mind that we have one system (the chain of particles) embedded into another system (the surrounding matrix, where here we present our results for one part of length $L$ of this infinitely extended matrix). 
When we perform a small affine deformation $\Delta L$ of the part of the matrix containing the chain, we alter the properties of the embedded system. 
In particular, the equilibrium distance $\ell$ between the embedded particles changes by a factor $(L + \Delta L)/L$. 
In our approach, this affects the pinning positions $x_i^{0} = i\ell$ of the particles as well as the spring constant $k_\textrm{pp} = \frac{3d}{8\ell} k_\textrm{mp}$, which is accounted for in the energetic change upon deformation. 
Furthermore, what we can calculate from this energetic change are only the contributions $\Delta p$ and $\Delta K$ of the embedded chain to the overall pressure and compression modulus of the composite. 
To obtain the overall pressure or compression modulus of the whole composite, the energetic change associated with the macroscopic deformation of the three-dimensional matrix would need to be included as well, which is beyond our particle-based approach. 

Figure~\ref{Fig.MC_vs_DFT_embedded_p_and_K} shows the contribution $\Delta p$ to the pressure as well as the contribution $\Delta K$ to the compression modulus as a function of the magnetic dipole moment. 
We can observe a linear decrease with $m^2$ in both quantities with good agreement between the DFT and MC calculations. 
The linear behavior is expected, because the magnetic interaction energy scales with $m^2$ and the particle chain remains relatively homogeneous while increasing the magnetic moment. 

\begin{figure}[b]
	\includegraphics[width=1.0\columnwidth]{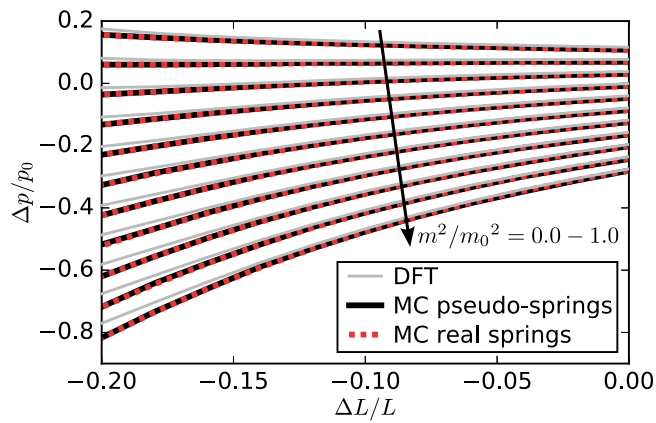}
	\caption{Pressure contribution $\Delta p$ as a function of an overall matrix compression $\Delta L$ for squared magnetic moments in the range of $m^2  = 0.0\,{m_0}^2 - 1.0\,{m_0}^2$ in steps of $0.1\,{m_0}^2$. These results constitute the contribution of the embedded chain to the overall stress-strain behavior in one part of the composite material. At vanishing magnetic moment, the pressure increases with the compression. Increasing the magnetic moment, however, leads to stronger magnetic attractions so that the pressure contribution decreases when the surrounding matrix is compressed. We find good agreement between DFT and MC results, especially for low compressions and magnetic moments. The deviations for high compressions and magnetic moments appear, presumably, for the reasons already described in Fig.~\ref{Fig.MC_vs_DFT_k4_mm}. Agreement between the pseudo-spring and real-spring MC is excellent under all conditions.}
	\label{Fig.MC_vs_DFT_embedded_stressstrain}
\end{figure}

As a final result, we present the contribution of the embedded chain to the stress-strain behavior of the composite material. 
For this purpose, we compress the surrounding matrix by $\Delta L$ and measure the pressure contribution of the embedded chain as a function of this compression. 
The results are shown in Fig.~\ref{Fig.MC_vs_DFT_embedded_stressstrain} for values of the squared magnetic moment in the range $m^2  = 0.0\,{m_0}^2 - 1.0\,{m_0}^2$. 
At vanishing magnetic moment, the pressure contribution slightly increases when compressing the system. 
This is  probably due to entropic effects that favor an elongated chain and, thus, work against a compression combined with a slight increase in the spring constant $k_\textrm{pp}$ in our description. 
Increasing $m^2$, however, leads to a stronger magnetic attraction between the particles. 
This renders an overall compression more favorable. 
Since decreasing the particle distance also enhances the magnetic attraction, we have a negative pressure contribution that increases in magnitude when compressing the system further.

Again, we can observe good agreement between DFT and MC. 
The best agreement is observed at small compressions and low magnetic moments. 
Remarkably, the pseudo-spring and the real-spring MC agree exceptionally well at all considered values of $\Delta L$ and $m^2$. 
This demonstrates, that our approach to map the spring network to effective interactions between indistinguishable particles is a promising approach also beyond the scope of our mean-field DFT. 

\section{Conclusions}
\label{Sec.Conclusions}

In summary, we have proposed a density functional theory to address ferrogel model systems, here evaluated in one spatial dimension. 
These systems are in principle non-liquid, because the particles are arrested by the elastic matrix surrounding them. 
To enable the investigation with statistical-mechanical theories, we map the elastic interactions onto effective pairwise interactions and, thus, make the particles indistinguishable. 

The one-dimensional nature of the ferrogel model systems investigated here poses a challenge, because thermal fluctuations have a special impact in one dimension. 
Fluctuations can become long-ranged and destroy periodic structural order. 
These fluctuations, driving the Landau-Peierls instability, are not resolved within our mean-field density functional theory. 
Therefore, within our first dipole-spring model we observe deviations from Monte-Carlo simulations where these fluctuations are included. 

In a second, more advanced approach, we explicitly model a linear particle chain embedded into a three-dimensional matrix. 
Within this model, the Landau-Peierls instability is counteracted by a stronger long-ranged coupling between the particles and a pinning potential that localizes the particles within the elastic matrix. 
Since the role of the fluctuations is therefore reduced, our density functional theory now provides results that are in good agreement with Monte-Carlo simulations. 
Numerous experimental realizations of such a systems exist \cite{collin2003frozen,varga2003smart,gunther2012xray, borbath2012xmuct,gundermann2013comparison,huang2016buckling,Gundermann2017_SmartMaterStruct}, see particularly the set-up in Ref.~\onlinecite{huang2016buckling}. 

For the future, it would be promising to extend the concept proposed here to higher spatial dimensions, that is to two-dimensional sheets of ferrogels or full three-dimensional samples. 
In those dimensions, the Landau-Peierls instability will be less relevant. 
We expect that especially for regular crystal-like particle arrangements, where the one-body density is regularly peaked, density functional theory is reliable and provides a useful framework to study the properties of these promising materials.
It will be challenging to extend the present analysis to include the dynamics of the colloidal particles by using the concept of dynamical density functional theory
\cite{Archer2004_JChemPhys,Marconi1999_JChemPhys,Espanol2009_JChemPhys,Loewen2017_incollection}. 
For particles of different sizes, or different dipole moments, the same ideas can be used to map the system onto binary and multicomponent systems.
Moreover, orientational degrees of freedom, such as rotating dipole moments or anisotropic particle shapes, can be tackled by density functional theory as well, both in statics \cite{Groh1996_PhysRevE,Klapp2000_JChemPhys} and dynamics \cite{Rex2007_PhysRevE,Menzel2016_JChemPhys}, and can therefore be treated within the same framework proposed here. 

Our results show that density functional theory can be used to describe non-liquid systems like ferrogels, still leading to reasonable results. 
More generally, we have established that mapping bead-spring models to effective potentials is a feasible approach to make them accessible to statistical-mechanical theories. 

These theories often make use of correlation functions as an input \cite{Denton1991_PhysRevA}, which here are related to the particle distribution in the ferrogel. 
Experimental extraction of the particle distribution and the corresponding correlation functions is still challenging \cite{Gundermann2017_SmartMaterStruct}. 
However, this route could be explored in the future once the available experimental techniques for particle detection in ferrogel materials are more advanced and particularly can address larger system sizes. 
These correlation functions could then help to construct effective pair potentials representing the real connectivity in the gel \cite{Hansen-Goos2006_EurophysLett}, providing a formal route for the mapping onto systems of indistinguishable particles. 

\appendix*
\section{Numerical relaxation scheme to obtain the equilibrium density profile from our DFT} 
Here, we describe in detail our numerical relaxation scheme to obtain the equilibrium density profile $\rho(x)$ minimizing $\Omega[\rho]$ within our DFT. 
Instead of directly solving Eq.~\eqref{Eq.DFT_Euler-Lagrange}, we perform a dynamical relaxation of the Lagrange functional
\begin{equation}
	\begin{aligned}
		\mathcal{L}[\alpha] &= \int_{0}^{L} \frac{1}{2} \dot{\alpha}(x)^2\, dx - \Omega[\alpha] \\
			&\qquad - \lambda \left( \int_{0}^{L} \exp\big(\alpha(x) \big) \,dx - \langle N \rangle \right) \, 
	\end{aligned}
	\label{Eq.DFT_grand_functional_dynamical}
\end{equation}
with respect to the logarithmic density profile $\alpha(x) = \ln\big( \rho(x) \big)$ \cite{Ohnesorge1994_PhysRevE}.
Minimizing with respect to the logarithmic density profile ensures that $\rho(x) = \exp\big( \alpha(x) \big)$ remains positive during the relaxation.
The artificial kinetic term $\int_{0}^{L} \frac{1}{2} \dot{\alpha}(x)^2 \, dx$ drives $\alpha(x)$  and thus $\rho(x)$ towards the minimum in the grand canonical free energy $\Omega[\rho]$. 
The Lagrange multiplier $\lambda$ with the corresponding constraint $\langle N \rangle = \int_{0}^{L} \exp\big( \alpha(x) \big) \, dx = \int_{0}^{L} \rho(x) \,dx$ allows us to set the average particle number $\langle N \rangle$ instead of the chemical potential $\mu$. 
This is more convenient for evaluating the pressure and the compression modulus defined as $p = -\frac{\partial \Omega}{\partial L}\big\vert_{\langle N \rangle, T}$ and $K = -L \frac{\partial p}{\partial L}\big\vert_{\langle N \rangle, T}$, respectively.

Solving the Euler-Lagrange equation
\begin{equation}
	\frac{d}{dt} \frac{\delta \mathcal{L}[\alpha]}{\delta \dot{\alpha}(x)} - \frac{\delta \mathcal{L}[\alpha]}{\delta \alpha(x)} \overset{!}{=} 0 \\
	\label{Eq.DFT_Euler-Lagrange_dynamical}
\end{equation}
then leads to the equation of motion for $\alpha(x)$
\begin{equation}
	\ddot{\alpha}(x) = -\rho(x) \left( \frac{\delta \Omega[\rho]}{\delta \rho(x)} + \lambda \right) \, .
	\label{Eq.DFT_equation_of_motion}
\end{equation}
The Lagrange multiplier $\lambda$ is determined by
\begin{equation}
	\begin{gathered}
		\frac{d^2}{dt^2} \int_{0}^{L} \exp\big( \alpha(x) \big) \, dx = 0 \quad\Leftrightarrow \\
		\int_{0}^{L} \rho(x) \left( \dot{\alpha}(x)^2 + \ddot{\alpha}(x) \right) \, dx = 0 \quad\Leftrightarrow \\
		\lambda = \int_{0}^{L} \rho(x) \left( \dot{\alpha}(x)^2 - \rho(x) \frac{\delta \Omega[\rho]}{\delta \rho(x)} \right) dx \, \bigg/  \int_{0}^{L}  \rho(x)^2 \, dx  \, .
	\end{gathered}
	\label{Eq.DFT_Lagrange_multiplier}
\end{equation}

In order to perform the numerical relaxation, we discretize the system into $n$ equally spaced sampling points such that one particle diameter $d$ is represented by $100$ points. 
The density profile, the potentials, and the radial distribution function are all defined on this numerical grid.
All integrals appearing in the calculation of the ``acceleration'' $\ddot{\alpha}(x)$ can then be solved numerically, making use of Fast-Fourier-Transforms in the case of convolution integrals.

We iterate the equation of motion for $\alpha(x)$ forward in time using the standard Velocity-Verlet scheme, obtaining the ``velocity'' $\dot{\alpha}(x)$ and an update for the density profile $\rho(x)$ in each time step $\Delta t$. 
To ensure that the constraint $\langle N \rangle = \int_{0}^{L} \rho(x) \,dx$ remains fulfilled, we renormalize $\rho(x)$ after each time step. 
The time step is variable and increases by a factor $1.1$ up to a maximum $\Delta t_\textrm{max} = 0.01$ when the grand canonical energy has decreased for $5$ consecutive time steps. The decrease in energy is monitored by the ``power'' $P = \int_{0}^{L} \dot{\alpha}(x) \ddot {\alpha}(x) \, dx$, which is supposed to be positive. If $P \leq 0$ occurs, we set $\dot{\alpha}(x) = 0$ and halve the time step. 
We consider the density profile $\rho(x)$ sufficiently close to equilibrium when our measure for the error $\varepsilon := \sqrt{\int_{0}^{L} \ddot{\alpha}(x)^2 \, dx}$ becomes smaller than $10^{-6}$. 
At that stage, the left hand side of Eq.~\eqref{Eq.DFT_equation_of_motion} as well as $\lambda$ are close to zero, so that we have $\frac{\delta \Omega[\rho]}{\delta \rho}(x) \approx 0$ as required by Eq.~\eqref{Eq.DFT_Euler-Lagrange}.

\begin{acknowledgments}
We thank Mate Puljiz for a stimulating discussion concerning the matrix-mediated interactions.
PC, AMM, and HL thank the Deutsche Forschungsgemeinschaft for support of this work through the priority program SPP 1681. 
\end{acknowledgments}

\bibliography{references}

\onecolumngrid
\clearpage
\begin{center}
\textbf{\large Supplemental material to: \\ A density functional approach to ferrogels}
\vspace{1.0\baselineskip}
\thispagestyle{empty}

\begin{minipage}{0.11\columnwidth}
\end{minipage}
\begin{minipage}{0.78\columnwidth}
\small

\indent In the main article, we have mapped our one-dimensional dipole-spring model containing magnetic particles that are distinguishable by their positions onto a description using pairwise interactions in terms of effective ``pseudo-springs'', such that the particles can be treated as indistinguishable.
Using this mapping, we have then compared results from density functional theory to those of Monte-Carlo simulations. 
Here, in a similar fashion, we compare results obtained from solving the Zerah-Hansen liquid-integral equation in comparison to those of Monte-Carlo simulations for various magnitudes of the magnetic moments of the particles. 
These results demonstrate, that other liquid-state theories using our pseudo-spring approximation lead to reasonable predictions as well. 
\end{minipage}
\begin{minipage}{0.11\columnwidth}
\end{minipage}
\vspace{1.0\baselineskip}

\end{center}

\twocolumngrid

\setcounter{equation}{0}
\setcounter{figure}{0}
\setcounter{table}{0}
\setcounter{page}{1}
\makeatletter
\renewcommand{\theequation}{S\arabic{equation}}
\renewcommand{\thefigure}{S\arabic{figure}}
\renewcommand{\bibnumfmt}[1]{[S#1]}
\renewcommand{\citenumfont}[1]{S#1}

Let us first describe the method of using the Zerah-Hansen (ZH) liquid-integral equation to obtain the equation of state, before we proceed to the comparison with Monte-Carlo (MC) simulations.
We use a numerical spectral method [S1-S3] to solve the Ornstein-Zernike equation [S4]
\begin{equation}\label{eq:O-Z}
\gamma(x) = \bar{\rho} \int\limits_{-\infty}^{\infty} dx' \big( \gamma(x') + c(x') \big) c(x-x')  
\end{equation}
for a bulk liquid of mean density $\bar{\rho}$. 
The functions $\gamma(x) = g(x) - 1 - c(x)$ and $c(x)$ are the indirect and direct correlation functions, respectively.
The latter is approximated here by the thermodynamically partially self-consistent Zerah-Hansen (ZH) closure [S5]
\begin{equation}\label{eq:ZH}
c(x) = \dfrac{e^{-\beta u_\textrm{r}(x)} \left[ f(x) - 1 + e^{f(x)\left(\gamma(x) -\beta u_\textrm{a}(x)\right)} \right]}{f(x)} -
       \gamma(x) - 1, 
\end{equation}
where $\beta = 1 / k_BT$ and the interaction potential $u(x) = u_\textrm{h}(x) + u_\textrm{m}(x) + \tilde{u}_\textrm{e}(x)$ 
between indistinguishable particles is split into the sum $u(x) = u_\textrm{r}(x) + u_\textrm{a}(x)$ of a repulsive part
\begin{equation}\label{eq:u_repulsive}
u_{r}(x) = \begin{cases}
	0	&	\textrm{for} \,\, x > x_\textrm{min}\, ,\\
	u(x) - u_\textrm{min}	&	\textrm{otherwise,}
\end{cases}
\end{equation}
and an attractive part
\begin{equation}\label{eq:u_attractive}
u_\textrm{a}(x) = \begin{cases}
	u(x)	&	\textrm{for} \,\, x > x_\textrm{min}\, ,\\
	u_\textrm{min}	&	\textrm{otherwise.}
\end{cases}
\end{equation}
Here, $u_\textrm{min} = u(x_\textrm{min})$ denotes the minimum of $u(x)$.
The mixing function $f(x) = 1 - e^{-\alpha x}$ depends on the non-negative mixing parameter $\alpha$
which is adjusted to achieve thermodynamic consistency with respect to the compression modulus:
At the numerically determined value of $\alpha$, the fluctuation-route expression [S4,S6]
\begin{equation}\label{eq:fluct_inv_compr}
K = \bar{\rho} k_B T \left( 1 - 2 ~ \bar{\rho} \int\limits_0^\infty c(x)~dx \right) 
\end{equation}
gives the same result as the virial-route expression [S4,S6]
\begin{equation}\label{eq:virial_inv_compr}
K = \bar{\rho} {\left. \dfrac{\partial p_\textrm{v}}{\partial \bar{\rho}} \right|}_{T},
\end{equation}
in which $p_\textrm{v}$ is the virial pressure.
The derivative in Eq.~\eqref{eq:virial_inv_compr} is numerically approximated by a finite difference 
and the virial pressure is calculated by numerical integration and solution of
\begin{equation}\label{eq:virial_pressure}
\dfrac{p_\textrm{v}}{\bar{\rho} k_B T}  = 1 + \bar{\rho} d g(d^{+}) - \bar{\rho} \beta \int\limits_{d}^{\infty} x  ~ \dfrac{d u(x)}{dx} ~ g(x) ~ dx. 
\end{equation}
Here, $g(d^{+}) = \lim_{x \to d} g(x > d)$ is the contact value of the radial distribution function. 

\begin{figure}
	\includegraphics[width=1.0\columnwidth]{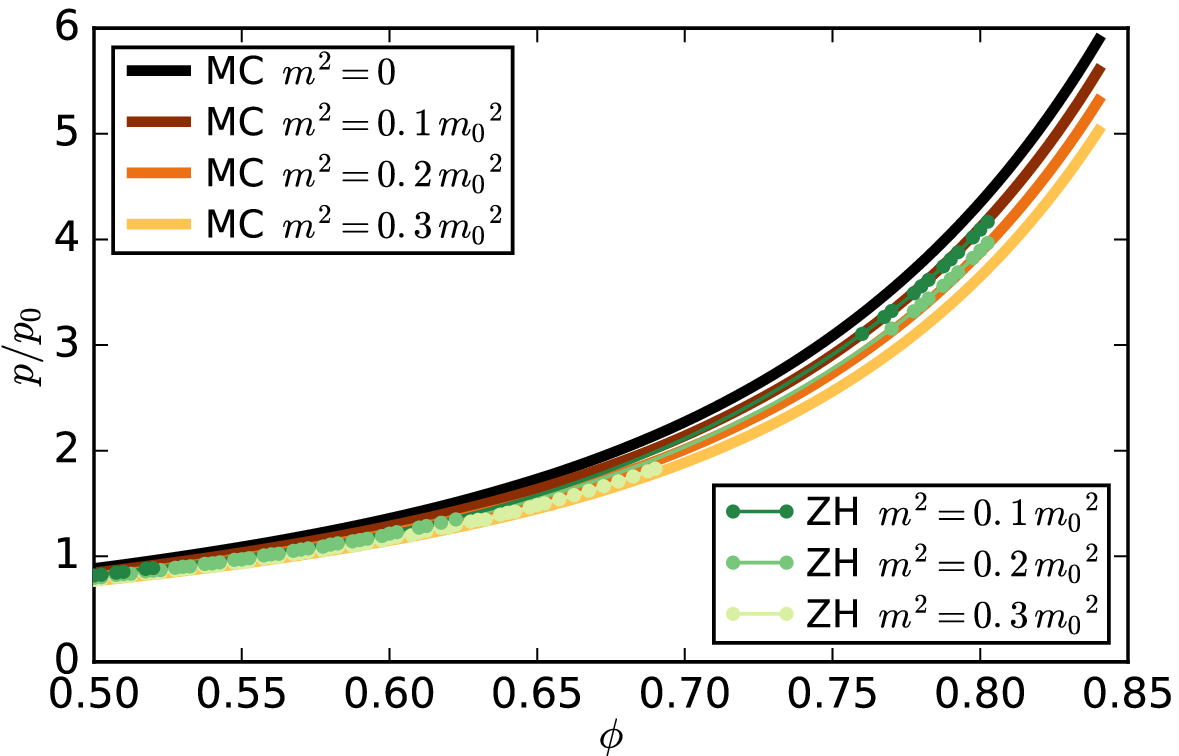}
	\caption{Equations of state obtained from bulk MC-simulation and ZH liquid-integral theory for various magnitudes of the magnetic moment. Thermodynamic consistency cannot be found for many parameter combinations, explaining the lack of ZH data points. Where it can be found, however, there is good agreement with the MC, especially at high packing fractions. Also the magnitude of the downward shift when increasing the magnetic moment seems to be accurately captured.}
	\label{Fig.MC_vs_ZH_k4_mm}
\end{figure}

We compare the equations of state obtained from this method with MC-simulations of a bulk pseudo-spring system.
For the latter we use a periodic box of width $L = 500d$ and otherwise proceed as for our finite systems in the main article. 
Figure~\ref{Fig.MC_vs_ZH_k4_mm} shows the equations of state obtained from both methods, ZH and MC, for various values of the magnetic moment.

Unfortunately, there are many parameter combinations for which the numerical ZH equation solver does not converge, because thermodynamic consistency with respect to the compression modulus is not found for any value of the mixing parameter $\alpha$. 
Nevertheless, for those parameters at which ZH solutions are available they agree remarkably well with the MC results. 
Especially at high packing fractions, the equation of state is accurately predicted by the ZH equation. 
Furthermore, the ZH solution captures correctly the decrease in pressure when the magnetic moment is increased. 
In contrast to density functional theory (DFT), the employed ZH equation is a theory for homogeneous, isotropic bulk liquids that is not expected to overestimate the tendency of the system to form regular structures and, in fact, does not even include the possibility of symmetry-breaking phase transitions. 
This might explain why the agreement between MC and ZH in Fig.~\ref{Fig.MC_vs_ZH_k4_mm} is better than between MC and DFT in Fig.~7 of the main article. 
However, the missing feature of reproducing localized density peaks at pre-set equilibrium positions, as required for the extended model in the main article, is an obvious drawback when compared to the DFT approach.

We have also solved the hypernetted chain (HNC) integral equation [S7] and the Percus-Yevick (PY) integral equation [S8] (results not shown). 
Both the HNC and the PY equations can be readily solved numerically in the complete parameter range of Fig.~\ref{Fig.MC_vs_ZH_k4_mm}. 
However, the (virial and fluctuation-route) equations of state predicted by the HNC and PY equation solution exhibit distinctive disagreement with our MC-simulation results. 
We conclude that thermodynamic consistency with respect to the compression modulus --- which is satisfied in the ZH equation solution, and which is lacking in both the HNC and the PY equations --- is a crucial feature.
In order to overcome the problem of missing solutions of the ZH equation in extended physical parameter ranges, it might be worthwhile to test alternative thermodynamically partially self-consistent closures of the Ornstein-Zernike equation in future work.


\end{document}